\documentstyle[twocolumn,aps,prl,epsf,floats]{revtex}
\newcommand{\beq}{\begin{equation}}
\newcommand{\eeq}{\end{equation}}
\newcommand{\bqa}{\begin{eqnarray}}
\newcommand{\eqa}{\end{eqnarray}}
\def\square{\vcenter{\vbox{\hrule height.4pt
          \hbox{\vrule width.4pt height8pt
          \kern8pt\vrule width.4pt}\hrule height.4pt}}}

\begin{document}
\twocolumn[\hsize\textwidth\columnwidth\hsize\csname
@twocolumnfalse\endcsname

%\noindent
%\hfill \mbox{BI-TP 97/28, CERN-TH/97-191, hep-ph/9708207}
%%\vspace*{4.8cm}
\vspace{-0.9cm}

\begin{flushleft}\hspace{12cm}
OHSTPY--HEP--T--98--015 \\ \hspace{12cm}
cond--mat/9808346 \\ \hspace{12cm}
%\today  \\
\end{flushleft}

\title{Critical Behaviour of a Homogeneous Bose 
Gas at Finite Temperature}
\author{Jens O. Andersen and Michael Strickland}
\address{Department of Physics, The Ohio State University, Columbus, OH 43210}
\date{\today}%April 7, 1997} %\today}
\maketitle

\begin{abstract}\noindent
A homogeneous non-relativistic Bose gas is investigated at finite temperature
using renormalization group methods. The  phase transition is shown
to be second order and the effective chemical potential and the 
effective quartic coupling vanish at the 
critical temperature. We obtain the critical exponent $\nu =0.73$ at leading
order in the derivative expansion.
\end{abstract}

\draft
\vspace{-0.2cm}
%\pacs{PACS numbers: 03.75.F, 05.70.F, 64.60A.}
\vskip1.5pc]

%\section{Introduction}

The remarkable achievement of Bose-Einstein condensation (BEC) of alkali
atoms in magnetic
traps~\cite{bec1} 
has created an enormous interest in the properties of dilute Bose gases.
A very recent review on trapped Bose gases can be found in~\cite{dalfovo}.
The homogeneous Bose gas at zero temperature 
was extensively studied in the fifties~\cite{Lee-Yang,Wu}. 
The properties of this system can be calculated in the loop expansion
which is an expansion in 
powers of $\sqrt{\rho a^3}$, where $\rho$ is the density
and $a$ is the S--wave scattering length. The one--loop correction to the
ground state energy was calculated in 1957 by Lee and Yang~\cite{Lee-Yang},
and 
the two--loop contribution was recently obtained by 
Braaten and Nieto~\cite{agus2}.
In Ref.~\cite{haug} Haugset {\it et al.} have 
studied this system at finite temperature, and 
included the daisy and superdaisy
diagrams by self--consistently solving a gap equation for the effective
chemical potential. The inclusion of these diagrams are essential
in order to satisfy the Goldstone theorem at finite temperature.
Biljsma and Stoof~\cite{henk} have applied the renormalization 
group~\cite{wilson}
to study the homogeneous Bose gas at finite temperature. 
Their results clearly demonstrate that the phase transition is second order
as expected, since this system is in the same universality class
as three--dimensional $xy$--model which is known to have a second order
phase transition.
Moreover, the critical temperature
increases by approximately 10$\%$ compared to the noninteracting Bose
gas~\cite{henk}. A review summarizing our current understanding
of homogeneous Bose gases can be found in~\cite{Shi-Griffin}.

In the present work we reconsider the non--relativistic homogeneous
spin zero Bose gas at finite
temperature. We focus in particular on the critical behaviour and
the calculation of critical exponents using RG techniques.
The work of Biljsma and Stoof~\cite{henk} only included the effects of two
and three--particle interactions by neglecting the running of higher
order vertices.
In our work we include the RG flow
of higher order particle interactions and show that the critical exponents
do not converge to the expected $O(2)$--model results
as successive terms are included.

The Bose gas can be described by an effective quantum field theory~\cite{e+g}.
As long as the momenta $p$ of the atoms are small compared to their 
inverse size, 
the interactions are effectively local and we can describe them
with a local quantum field theory. Since we assume that all the atoms
have the same spin we can describe them by a complex spin zero 
field:
\bqa
\psi={1\over \sqrt{2}}\left[\psi_1+i\psi_2\right].
\eqa 
The symmetries are Galilean invariance and a global $O(2)$--symmetry.
The Euclidean Lagrangian reads
\bqa\nonumber
{\cal L}_E&=&\psi^{\dagger}\partial_{\tau}\psi+\frac{1}{2m}\nabla\psi^{\dagger}
\cdot\nabla\psi-\mu\psi^{\dagger}\psi
\\ 
\label{l}
&&
+g(\psi^{\dagger}\psi)^2+\ldots.
%+V(\psi^{\dagger}\psi).
\eqa
Here, $\mu$ is the chemical potential. 
The ellipses indicate all operators that are higher order in the number of
fields $\psi$
and derivatives and  
satisfy the
$O(2)$--symmetry. The interaction $g(\psi^{\dagger}\psi)^2$
represents $2\rightarrow 2$ scattering and the coupling 
constant $g$ is proportional to the $S$--wave scattering length $a$:
\beq
g=\frac{2\pi a}{m}.
\eeq
In the following we consider the dilute gas $\rho a^3\ll 1$, which implies
that we only need to retain the quartic interaction in the bare Lagrangian 
Eq.~(\ref{l})~\cite{henk}. 
We also set $2m=1$.

In a field theoretic language, BEC is described
as spontaneous symmetry breaking of the $O(2)$--symmetry and the complex
field 
$\psi$ acquires a nonzero vacuum expectation value $v$.
%\bqa
%\psi={1\over \sqrt{2}}\left[\psi_1+i\psi_2\right],\hspace{0.7cm}
%\psi_1\rightarrow v+\psi_1,\hspace{0.7cm}\psi_2\rightarrow \psi_2.
%\eqa
The propagator is a $2\times 2$ matrix
\bqa
\Delta (\omega_{n},{p})&=&\frac{1}{\omega_{{p}}^2
+\omega_{n}^2}\left(\begin{array}{cc}
\epsilon_{{p}}+V^{\prime}&-\omega_{n} \\
\omega_{n}&\epsilon_{{p}}+V^{\prime}+V^{\prime\prime}v^2
\end{array}\right), 
\eqa
where
\bqa \nonumber
V&=&-{1\over2}\mu v^2+{g\over 4}v^4\\ \nonumber
\epsilon_{{p}}&=&p^2\\ \nonumber
\omega_n&=&2\pi nT\\
\omega_{p}&=&
\sqrt{\left[\epsilon_{p}+V^{\prime}(v)+V^{\prime\prime}(v)v^2
\right]\left[\epsilon_{p}+V^{\prime}(v)\right]},
\eqa
and primes denote differentiation with respect to $v^2/2$.

The grand canonical partition function has a path integral representation
\bqa
Z=\int{\cal D}\psi^{\dagger}{\cal D}\psi e^{-\int_0^{\beta}d\tau\int d^dx
{\cal L}_E}.
\eqa
In order to implement the renormalization group we modify the action by adding
a piece containing a cutoff function, $R_k^{\Lambda}(p)$, to the action
\bqa
S_{\beta,k}[\psi,\psi^{\dagger}]=\int_{0}^{\beta}d\tau\int d^dx
\Big\{R_k^{\Lambda}
\nabla\psi^{\dagger}\cdot 
\nabla\psi+S[\psi,\psi^{\dagger}]\Big\},
\eqa 
where
\bqa
S[\psi,\psi^{\dagger}]=\int_{0}^{\beta}d\tau\int d^dx{\cal L }_E.
\eqa
The function $R_k^{\Lambda}(p)$
suppresses all modes
in the path integral with momenta $p$
less than $k$ and larger than $\Lambda$, where $k$ and $\Lambda$ are
our infrared and ultraviolet cutoffs, respectively. In the present work
we choose a sharp cutoff such that 
\bqa
R_k^{\Lambda}(p)=\left\{\begin{array}{c}
0,\hspace{1cm}k<p<\Lambda,\\
\infty,\hspace{1cm}\mbox{otherwise}.
\end{array}\right.
\eqa
The flow equation or 
renormalization group equation for the effective action 
$\Gamma_{\beta,k}[v]$ is obtained by taking a derivative with respect to
the infrared cutoff $k$~[11--13]. This equation specifies how 
$\Gamma_{\beta,k}[v]$ changes as $k$ is lowered:
\bqa\nonumber
{\partial\Gamma_{\beta,k}\over\partial k}&=&{1\over 2}T\sum_n\int 
{d^dp\over(2\pi)^d}
\left({{\partial R_k^{\Lambda}}\over\partial k}\right)\\
&&\times\mbox{Tr}\left[R_k^{\Lambda}\delta_{ab}
+{\partial^2\Gamma_{\beta,k}\over\partial v_a\partial
v_b}
\right]^{-1}.
\eqa
The sum is over the Matsubara frequencies which take on the values
$2\pi n T$ for bosons and the integration is over $d$--dimensional 
momentum space.
The trace is over internal indices.
%Furthermore, $mbox{Tr}$ denotes the trace of the 
Also note that the flow equation explicitly depends on the cutoff function 
$R_k^{\Lambda}$ (see also~\cite{mike3}).

The effective action can be expanded in powers of derivatives:
\bqa\nonumber
\Gamma_{\beta,k}[v]&=&\int_0^{\beta}d\tau\int d^dx
\Big\{U_{\beta,k}[v]
+Z^{(1)}_{\beta,k}[v]\epsilon_{ij}v_i\partial_{\tau}v_j \\
&&+Z^{(2)}_{\beta,k}[v](\nabla v_i)^2
+\ldots.\Big\}
\eqa
To leading order in the derivative expansion the effective action 
$\Gamma_{\beta,k}$
is (up to a volume factor) simply
the effective potential $U_{\beta,k}$, and $Z^{(1)}_{\beta,k}$ 
and $Z^{(2)}_{\beta,k}$
are both equal to unity. 
The RG--equation for the effective
potential reads
\bqa
\label{rg1}
k\frac{\partial U_{\beta,k}}{\partial k}=
-\frac{S_dk^d}{2}\omega_{{k}}-S_dk^dT
\ln\left[1-e^{-\beta\omega_{{k}}}\right],
\eqa
where
\bqa
%\epsilon_k&=&k^2,\\
\omega_{{k}}(v)&=&\sqrt{\left[\epsilon_{k}+U^{\prime}_{\beta,k}
+U^{\prime\prime}_{\beta,k}v^2\right]\left[\epsilon_{{k}}+U^{\prime}_{\beta,k}\right]}\\
S_d&=&{2\over(4\pi)^{d/2}\Gamma (d/2)}.
\eqa
This equation will be derived in~\cite{expand}.
The first term on the right hand side of Eq.~(\ref{rg1}) is the $T=0$ contribution to the
RG--equation and includes the quantum fluctuations. 
This equation interpolates between the bare theory for $k=\Lambda$
and the physical theory at $T\neq 0$ for $k=0$, since we integrate out
both quantum and thermal modes as we lower the cutoff.
This implies that the boundary condition for the RG--equation is the
{\it bare} potential.
In Ref.~\cite{pepperoni} renormalization group ideas have been applied
to relativistic $\lambda\phi^4$--theory using the real time formalism.
Here, one can separate the propagator into a
quantum and a thermal part, and
the infrared cutoff is imposed only on the 
thermal part of the propagator.
This implies that the theory interpolates between the physical theory
at $T=0$ and the physical theory at $T\neq 0$. Hence, the boundary condition
of the RG--equation is the physical effective potential at 
$T=0$~\cite{pepperoni}.
%In the following we subtract off the zero temperature term in Eq.~(\ref{rg1}) above and
%consider the latter approach.
%Thus our renormalization group equation reads
%\bqa
%\label{rg2}
%k\frac{\partial U_{\beta,k}}{\partial k}=
%-S_dk^dT
%\ln\left[1-e^{-\beta\omega_{{k}}}\right],
%\eqa

\begin{figure}[htb]
%\vspace*{-4.0cm}

\hspace{1cm}
\epsfysize=4cm
\centerline{\epsffile{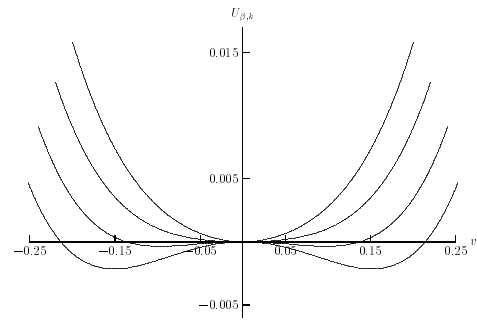}}
\\
%\vspace*{-4.6cm}
\caption[a]{The RG--improved effective potential $U_{\beta,k}[v]$
for different values of the
temperature. The phase transition is clearly second order.}
\label{pot}
\end{figure}

We have solved Eq.~(\ref{rg1}) for $d=3$ 
numerically for different values of the 
temperature and the result is displayed in Fig.~\ref{pot}.
The curves clearly show
a second order phase transition, which is expected
from universality arguments. 

\begin{figure}[htb]

\vspace*{-4.0cm}

\hspace{1cm}
\epsfysize=13cm
\centerline{\epsffile{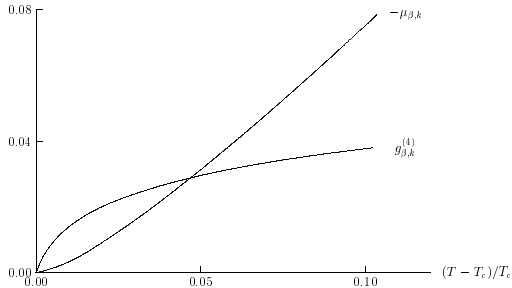}}
\\
\vspace*{-4.6cm}
\caption[a]{The effective chemical potential $\mu_{\beta,k}$ and 
the effective quartic coupling $g^{(4)}_{\beta,k}$ 
near the critical temperature. Both vanish at $T_c$.}
\label{g2}
\end{figure}

The effective chemical potential $\mu_{\beta, k}$
as well as the coupling constant
$g_{\beta, k}^{(4)}$ 
(defined as the discrete first and second derivatives of the
effective potential with respect to $v^2/2$) are displayed in Fig.~\ref{g2}. 
We see that both quantities vanish at the
critical point. Moreover, $g_{\bar{\beta}, k}^{(6)}$
goes to a non-zero constant at $T_c$. 
%This reflects that the polynomial expansion breaks
%down near the critical point. This breakdown has also been 
%demonstrated in the case of relat(ivistic $\lambda\phi^4$ theory
%in Ref.~\cite{mike2}. 
The inclusion of wavefunction renormalization effects
turns the marginal operator 
$g_{\bar{\beta}, k}^{(6)}$ into an irrelevant operator that diverges
at the critical temperature~\cite{mike2}.

In order to locate the fixed points we write the effective potential in
dimensionless form and make a series expansion around the origin:
\bqa\nonumber
\bar{U}_{\bar{\beta},k}&=&\beta k^{-d}U_{\beta,k}\\ \nonumber
\bar{v}&=&\beta^{1/2}k^{(2-d)/2}v\\\nonumber
\bar{\omega}_k&=&k^{-2}\omega_k\\\nonumber
\bar{\beta}&=&\beta k^2\\\nonumber
\bar{\mu}_{\bar{\beta},k}&=&k^{-2}\mu_{\beta,k}\\\nonumber
\bar{g}^{(4)}_{\bar{\beta},k}&=&\beta^{-1}k^{d-4}g^{(4)}_{\beta,k}\\\nonumber
%\bar{g}^{(6)}_{\bar{\beta},k}&=&\ldots.\\\nonumber
\vdots&=&\vdots\\
\bar{U}_{\bar{\beta} ,k}(\bar{v})&=&\sum_{n=1}^{\infty}
\frac{\bar{g}^{(2n)}_{\bar{\beta}, k}}{n!}\left({\bar{v}^{2}\over 2}\right)^n,
\eqa
and $\bar{\mu}_{\bar{\beta},k}=-\bar{g}^{(2)}_{\bar{\beta},k}$.
Truncating the series after two terms, we obtain the following set of
equations
\bqa%\nonumber
\label{1l}
k\frac{\partial \bar{\mu}_{\bar{\beta} ,k}}{\partial k}&=&
-2\bar{\mu}_{\bar{\beta}}+S_d\bar{\beta}
\bar{g}^{(4)}_{\bar{\beta} ,k}
\coth[\bar{\beta} (1-\bar{\mu}_{\bar{\beta} ,{k}})/2]\\ \nonumber
%\frac{1}{e^{\bar{\beta} (1-\bar{\mu}_{\bar{\beta} ,{k}})}-1}
\label{d}k\frac{\partial \bar{g}^{(4)}_{\bar{\beta} ,k}}{\partial k}
&=&(d-4)\bar{g}^{(4)}_{\bar{\beta},k}+
S_d\bar{\beta}[\bar{g}^{(4)}_{\bar{\beta} ,k}]^2\\ \nonumber
&&\times
\left[\frac{1}{2(1-\bar{\mu}_{\bar{\beta},{k}})}
\coth[\bar{\beta} (1-\bar{\mu}_{\bar{\beta} ,{k}})/2]
%(e^{\bar{\beta} (1-\bar{\mu}_{\bar{\beta} ,{k}})}-1)}
\right. 
\\% \nonumber
&&
\left.+\frac{4\bar{\beta} e^{\bar{\beta} (1-\bar{\mu}_{\bar{\beta} ,{k}})}}
{(e^{\bar{\beta} (1-\bar{\mu}_{\bar{\beta} ,{k}})}-1)^2}\right].
\eqa
A similar set of equations has also been obtained
in Ref.~\cite{henk} by considering the one--loop diagrams that contribute
to the running of the different vertices (They use the operator formalism and
normal ordering so the zero temperature part of the tadpole vanishes).

Expanding in powers of $\bar{\beta} (1-\bar{\mu}_{\bar{\beta} ,{k}})$
and introducing the variables 
\beq
r={\bar{\mu}_{\bar{\beta},k}\over 1-\bar{\mu}_{\bar{\beta},k}},\hspace{1cm}
s={\bar{g}^{(4)}_{\bar{\beta},k}\over (1-\bar{\mu}_{\bar{\beta},k})^2},
\eeq
the RG--equations can be written as
\bqa
\frac{\partial r}{\partial k}&=&-2\left[1+r\right]
\left[r-S_ds\right]\\
\frac{\partial s}{\partial k}&=&-s\left[\epsilon+4r-9S_ds\right].
\eqa
Here, $\epsilon=4-d$.
We have the trivial Gaussian fixed point $(0,0)$ as well as the
infinite temperature Gaussian fixed point $(-1,0)$. Finally, for 
$\epsilon>0$ there is the
infrared Wilson--Fisher fixed point 
$\left(\epsilon/5,\epsilon/\left(5S_d\right)\right)$~\cite{wilson}.

One can now calulate the critical exponent $\nu$ which is related
to the correlation length $\xi$
through
\bqa
\xi\sim |T-T_c|^{-\nu}.
\eqa
Linearizing around the fixed point, we find the eigenvalues
$(-1.278,1.878)$. This implies that the critical exponent is $\nu=0.532$ in
agreement with the result of Biljsma and Stoof~\cite{henk}.

\begin{figure}[htb]

%\vspace*{-4.2cm}

\hspace{1cm}
\epsfysize=4cm
\centerline{\epsffile{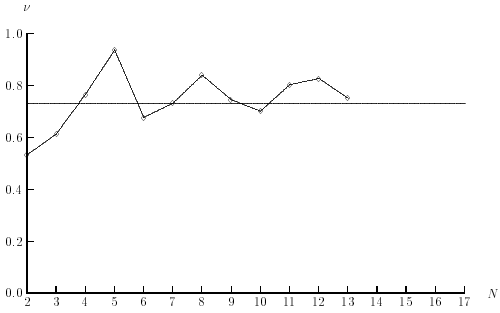}}
%\vspace*{-4.8cm}
\\
\caption[a]{The critical exponent $\nu$ as a function of number of terms $N$ 
in the polynomial expansion.}
\label{terms}
\end{figure}

In order to check the convergence of this expansion, we 
have repeated the calculation including more terms in the
Taylor expansion of the effective potential. The critical exponent $\nu$
as a function of the number of terms $N$ is shown in Fig.~\ref{terms}.
The critical exponent $\nu$ fluctuates around the value $0.73$ and should
be compared to experiment ($^{4}$He) and the $\epsilon$--expansion which
both give a value of $0.67$~\cite{zinn}. Moreover, our result agrees with
that of Morris, who considered the relativistic $O(2)$--model
in $3d$ at zero temperature~\cite{morris}. 
The fact that the critical exponent oscillates as a function of $N$ reflects
that the polynomial expansion of the effective potential breaks down
near the critical temperature~\cite{mike2,morris,japs}.

To improve our results for the critical
exponents, we must go beyond leading order in the derivative expansion.

We obtain the same critical behaviour as in the three--dimensional
$O(2)$--theory as expected from universality. This can 
be seen directly, without computation, by considering the dimensionless form of
Eq.~(\ref{rg1}):
\bqa\nonumber
0&=&
\left[k{\partial\over \partial k}-{1\over 2}(d-2)\bar{v}\partial_{\bar{v}}
+d\right]\bar{U}_{\bar{\beta},k}\\
&&+{S_d\over 2}\bar{\beta}\bar{\omega}_k
+S_d\ln\left[
1-e^{-\bar{\beta}\bar{\omega}_k}\right].
\eqa
The critical 
potential is now found by demanding that~\cite{morris}
\bqa
k{{\partial \bar{U}_{\bar{\beta},k}}\over{\partial k}}=0,
\eqa
and then expanding in powers of $\bar{\beta}\bar{\omega}_k$:
\bqa
\left[-{1\over 2}(d-2)\bar{v}\partial_{\bar{v}}
+d\right]\bar{U}_{\bar{\beta},k}=
-{S_d\over 2}\bar{\beta}\bar{\omega}_k
-S_d\ln
\left[\bar{\beta}\bar{\omega}_k\right].
\eqa
Taking the limit $\bar{\beta}\rightarrow 0$~\cite{mike2} and
ignoring the piece which is independent of $v$, this leads to
\bqa\nonumber
\left[-{1\over 2}(d-2)\bar{v}\partial_{\bar{v}}
+d\right]\bar{U}_{\bar{\beta},k}&=&-{S_d\over2}
\Bigg[\ln\left[1+\bar{U}^{\prime}\right] \\
&&+\ln\left[1+\bar{U}^{\prime}+\bar{U}^{\prime\prime}\bar{v}^2\right]\Bigg].
\eqa
This is exactly the same equation as obtained by Morris 
for an $O(2)$--symmetric scalar theory in three dimensions
to leading order in the derivative
expansion~\cite{morris}.
Therefore, the results for the 
critical behaviour at leading order in the derivative
expansion will be the same as those obtained in the three-dimensional
$O(2)$--model 
at zero temperature.

In the present letter we have explicitly
demonstrated that the phase transition of the
homogeneous non--relativistic Bose gas is second order and that the critical
behaviour is the same as in the three--dimensional scalar $O(2)$--model. 
Both properties are
expected from universality arguments. 

One natural extension of the present work is to 
include wave function renormalization effects by going
to second order in the derivative expansion and investigate 
noncritical quantities such as the critical temperature
and the superfluid fraction in the system.

This work was supported in part by the U.~S. Department of Energy,
Division of High Energy Physics, under Grant DE-FG02-91-ER40690, by
the National Science Foundation under Grants No. PHY--9511923 and PHY--9258270,
and by a Faculty Development Grant from the Physics Department of The 
Ohio State University.
J. O. A. was also supported in part 
by a Fellowship 
from the Norwegian Research Council (project 124282/410).
The authors would like to thank the organizers of the ``5th International 
Workshop on Thermal Field Theory and their Applications'' for a stimulating
meeting.
\vspace{-0.6cm}
%\appendix
 
\end{document}